\documentclass[
    ,final            
  sort&compress
  ]
  {aipproc}

\layoutstyle{6x9}

\newcommand{\beq}{\begin{equation}}
\newcommand{\eeq}{\end{equation}}
\newcommand{\unity}{1\hspace{-0.15cm}1}

\begin{document}

\title{Bimaximal neutrino mixing and weak complementarity with $\mathbf{S_4}$ discrete symmetry}


\author{Luca Merlo}{
  address={Dipartimento di Fisica `G.~Galilei', Universit\`a di Padova \&
INFN, Sezione di Padova,\\ Via Marzolo~8, I-35131 Padua, Italy -- merlo@pd.infn.it}
}

\begin{abstract}
The neutrino oscillation data are well explained by the tri-bimaximal pattern. Recently a paper appeared showing that also the bimaximal pattern could be a very good starting point in order to describe the lepton mixing. In this paper I review both the flavour structures and then I present an explicit model.
\end{abstract}

\maketitle


\section{The lepton mixing matrix}

The solar and the atmospheric neutrino anomalies are well explained by the oscillation of three massive neutrinos. The pattern of the mixings is characterized by two large angles and a small one \cite{Data}: the atmospheric angle $\theta_{23}$ is compatible with a maximal value, but the accuracy admits relatively large deviations; the solar angle $\theta_{12}$ is large, but about $5\sigma$ errors far from the maximal value; the reactor angle $\theta_{13}$ only has an upper bound, which can be parameterized at $3\sigma$ as $\sin\theta_{13}\leq\lambda_C$, where $\lambda_C\simeq0.23$ is the Cabibbo angle. Measuring the reactor angle in future experiments represents a fundamental task in order to understand the pattern of the neutrino mixing.

\subsection{The tri-bimaximal mixing}

Within measurement errors, the observed neutrino mixing matrix is compatible with the tri-bimaximal (TB) form \cite{TB}. The best measured neutrino mixing angle $\theta_{12}$ is just about 1$\sigma$ below the TB value, while the other two angles are well inside the 1$\sigma$ interval. The most generic mass matrix of the TB-type is given by
\beq
m_\nu^{TB}=\left(
  \begin{array}{ccc}
    x & y & y \\
    y & z & x+y-z \\
    y & x+y-z & z \\
  \end{array}
\right)\;.
\eeq
This matrix satisfies to the $\mu-\tau$ symmetry and to the so-called magic symmetry, for which $(m_\nu^{TB})_{1,1}+(m_\nu^{TB})_{1,3}=(m_\nu^{TB})_{2,2}+(m_\nu^{TB})_{2,3}$. It is diagonalized by a unitary transformation in such a way that $(m_\nu^{TB})_{\mathit{diag}}=(U^{TB})^T m_\nu^{TB}U^{TB}$, where the unitary matrix is given by
\beq
U^{TB}=\left(
         \begin{array}{ccc}
           \sqrt{2/3} & 1/\sqrt3 & 0 \\
           -1/\sqrt6 & 1/\sqrt3 & -1/\sqrt2 \\
           -1/\sqrt6 & 1/\sqrt3 & +1/\sqrt2 \\
         \end{array}
       \right)\;.
\eeq
Note that $U^{TB}$ does not depend on the mass eigenvalues, in contrast with the quark sector, where the entries of the CKM matrix can be written in terms of the ratio of the quark masses. Moreover it is a completely real matrix, since the factors with the Dirac phase vanish (the Majorana phases can be factorized outside).

In a series of papers \cite{A4} it has been pointed out that a broken flavour symmetry based on the discrete group $A_4$ appears to be particularly suitable to reproduce this specific mixing pattern as a first approximation. Other solutions based on alternative discrete or continuous flavour groups have also been considered \cite{FHLM_Tp,BMM_S4,Continuous}, but the $A_4$ models have a very economical and attractive structure, e.g. in terms of group representations and of field content. In all these models, when the symmetry is broken, some corrections to the mixing angles are introduced: in general all of them are of the order of $\lambda_C^2$ and therefore these models indicate a value for the reactor angle which is well compatible with zero (for a different approach see \cite{Lin_Reactor}).

\subsection{The bimaximal mixing}

There is an experimental hint for a non-vanishing reactor angle \cite{Fogli} and, if a value close to the present upper bound is found in the future experiments, this could be interpreted as an indication that the agreement with the TB mixing is only accidental. Looking for an alternative leading principle, it is interesting to note that the data suggest a numerical relationship between the lepton and the quark sectors, known as the complementarity relation, for which $\theta_{12}+\lambda_C\simeq\pi/4$. However, there is no compelling model which manages to get this nice feature, without parameter fixing. Our proposal is to relax this relationship. Noting that $\sqrt{m_\mu/m_\tau}\simeq\lambda_C$, we can write the following expression, which we call weak complementarity relation
\beq
\theta_{12}\simeq\frac{\pi}{4}-\mathcal{O}\left(\sqrt{\frac{m_\mu}{m_\tau}}\right)\;.
\eeq
The idea is first to get a maximal value both for the solar and the atmospheric angles and then to correct $\theta_{12}$ with relatively large terms. To reach this task, the bimaximal (BM) pattern can be extremely useful: it corresponds to the requirement that $\theta_{13}=0$ and $\theta_{23}=\theta_{12}=\pi/4$. The most general mass matrix of the BM-type and its diagonalizing unitary matrix are then given by
\beq
m_\nu^{BM}=\left(
  \begin{array}{ccc}
    x & y & y \\
    y & z & x-z \\
    y & x-z & z \\
  \end{array}
\right)\;,\qquad\quad
U^{BM}=\left(
         \begin{array}{ccc}
           1/\sqrt2 & -1/\sqrt2 & 0 \\
           1/2 & 1/2 & -1/\sqrt2 \\
           1/2 & 1/2 & +1/\sqrt2 \\
         \end{array}
       \right)\;,
\eeq
where $m_\nu^{BM}$ satisfies to the $\mu-\tau$ symmetry and to an additional symmetry for which $(m_\nu^{BM})_{1,1}=(m_\nu^{BM})_{2,2}+(m_\nu^{BM})_{2,3}$. Note that $U^{BM}$ does not depend on the mass eigenvalues and is completely real, like the TB pattern.

Starting from the BM scheme, the corrections introduced from the symmetry breaking must have a precise pattern: $\delta\sin^2\theta_{12}\simeq\lambda_C$, while $\delta\sin^2\theta_{23}\leq\lambda_C^2$ and $\delta\sin\theta_{13}\leq\lambda_C$ in order to be in agreement with the experimental data. This feature is not trivially achievable.


\section{The model building}
In this part I present a flavour model in which the neutrino mixing matrix at the leading order (LO) is the BM scheme, while the charged lepton mass matrix is diagonal with hierarchical entries; moreover the model allows for corrections that bring the mixing angles in agreement with the data (for details see \cite{AFM_Bimax}). The strategy is to use the flavour group $G_f=S_4\times Z_4\times U(1)_{FN}$, where $S_4$ is the group of the permutations of four objects, and to let the SM fields transform non-trivially under $G_f$; moreover some new fields, the flavons, are introduced which are scalars under the SM gauge symmetry, but transform under $G_f$; these flavons, getting non-vanishing vacuum expectation values (VEVs), spontaneously break the symmetry in such a way that two subgroups are preserved, $G_\ell=Z_4$ in the charged lepton sector and $G_\nu=Z_2\times Z_2$ in the neutrino sector. It is this breaking chain of $S_4$ which assures that the LO neutrino mixing matrix, $U_\nu$, is the BM pattern and that the charged lepton mass matrix is diagonal (and therefore $U_\ell=\unity$). The additional terms $Z_4\times U(1)_{FN}$ forbid dangerous operators and allow for the correct charged lepton mass hierarchy. From the definition of the lepton mixing matrix $U\equiv (U_\ell)^\dag U_\nu$, the only which can be observable, it follows that $U$ coincides with $U^{BM}$.

\begin{figure}[h!]
 \centering
   {\includegraphics[width=11cm]{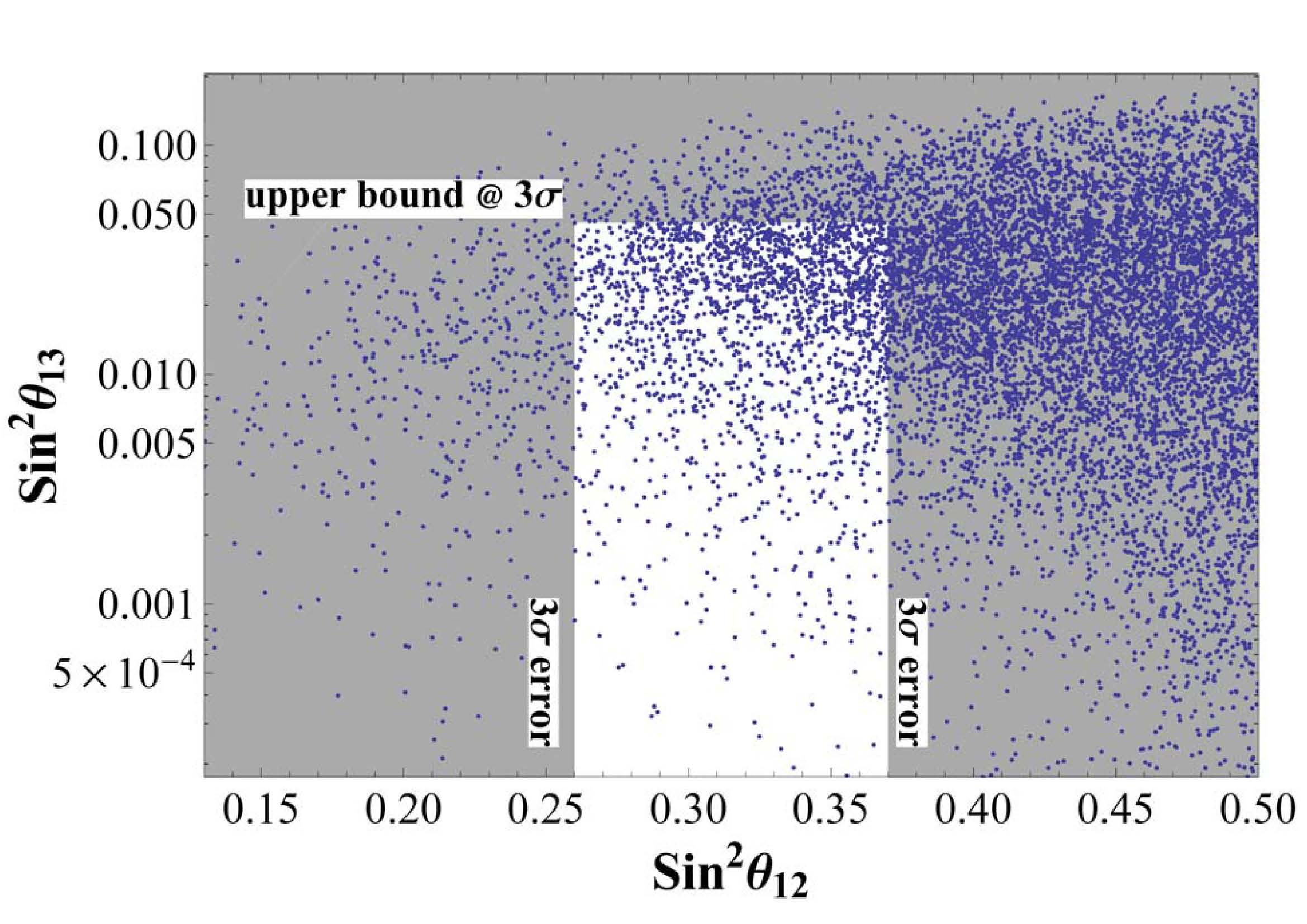}}
 \caption{$\sin^2\theta_{13}$ as a function of $\sin^2\theta_{12}$ is plotted, following eqs. (\ref{sinNLO}). The plot is symmetric with respect $\sin^2\theta_{12}=0.5$ and we report here only the left-part. The parameters $a$ and $b$ are treated as random complex numbers of absolute value between 0 and 0.30. The gray bands represents the regions excluded by the experimental data \cite{Data}: the horizontal one corresponds to the $3\sigma$-upper bound for $\sin^2\theta_{13}$ of 0.46 and the vertical ones to the region outside the $3\sigma$ error range $[0.26 - 0.37]$ for $\sin^2\theta_{12}$.}
 \label{fig0}
\end{figure}

Thanks to the particular VEV alignment, the next-to-leading order (NLO) corrections, coming from the higher order terms, are not democratic and the corrected mixings are
\beq
\sin^2\theta_{12}=\frac{1}{2}-\frac{1}{\sqrt2}(a+b)\;,\qquad
\sin^2\theta_{23}=\frac{1}{2}\;,\qquad
\sin\theta_{13}=\frac{1}{\sqrt2}(a-b)\;,
\label{sinNLO}
\eeq
where $a$ and $b$ parameterize the VEVs of the flavons. When $a$ and $b$ are of the order of the Cabibbo angle, then $\theta_{12}$ is brought in agreement with the experimental data; in the meantime the reactor angle is corrected of the same amount, suggesting a value for $\theta_{13}$ close to its present upper bound. Note that the atmospheric angle remains uncorrected at this order.
Any quantitative estimates are clearly affected by large uncertainties due to the presence of unknown parameters of order one, as we can see in figure \ref{fig0}, but in our model a value of $\theta_{13}$ much smaller than the present upper bound would be unnatural.
\\

It is then interesting to verify the agreement of the model with other sectors of the neutrino physics, such as the $0\nu2\beta$-decay and the leptogenesis (See \cite{ABMMM_Lepto} for a general approach). The result of the analysis is that the model presents a normal ordered -- moderate hierarchical or quasi degenerate -- spectrum with a suggested lower bound for the lightest neutrino mass and for the effective $0\nu2\beta$-mass parameter $|m_{ee}|$ of about 0.1 meV. On the other hand it is compatible with the constraints from leptogenesis as an explanation of the baryon asymmetry in the Universe.

\begin{theacknowledgments}
I thank the organizers of \emph{SUSY 2009 -- 17th International Conference on Supersymmetry and the Unification of Fundamental Interactions} for giving the opportunity to present my talk and for the kind hospitality in Boston.
Alike I thank Guido Altarelli and Ferruccio Feruglio for the pleasant and advantageous collaboration.
\end{theacknowledgments}

\vspace{-0.5cm}

\bibliographystyle{aipproc}

\end{document}